# Effect of Personality Traits on UX Evaluation Metrics: A Study on Usability Issues, Valence-Arousal and Skin Conductance

**Alexandros Liapis**
School of Science & Technology, Hellenic Open University, Patras, Greece
aliapis@eap.gr

**Christos Katsanos**
Department of Informatics, Aristotle University of Thessaloniki, Thessaloniki, Greece
ckatsanos@csd.auth.gr

**Michalis Xenos**
Department of Computer Engineering & Informatics, University of Patras, Patras, Greece
xenos@ceid.upatras.gr

**Theofanis Orphanoudakis**
School of Science & Technology, Hellenic Open University, Patras, Greece
fanis@eap.gr

## ABSTRACT

Personality affects the way someone feels or acts. This paper examines the effect of personality traits, as operationalized by the Big-five questionnaire, on the number, type and severity of identified usability issues, physiological signals (skin conductance), and subjective emotional ratings (valence-arousal). Twenty-four users interacted with a web service known, from a previous study, to have usability issues and then participated in a retrospective thinking aloud session. Results revealed that the number of usability issues is significantly affected by the Openness trait.

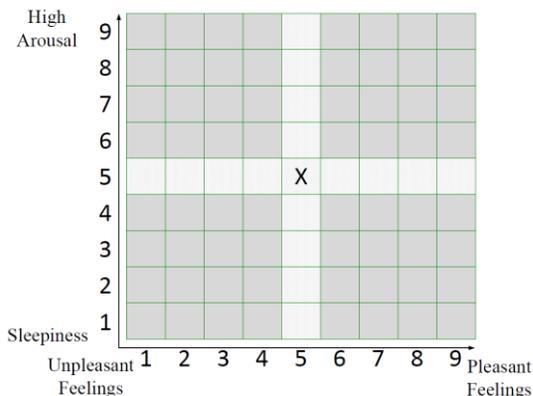

**Figure 1: The Affect Grid tool. If someone feels neutral, then the middle cell (5,5) of the matrix is expected to be selected.**

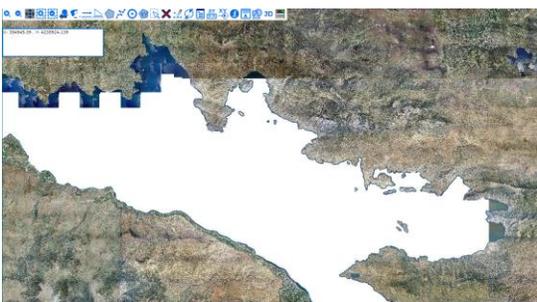

**Figure 2: The system evaluated in this study was the National Cadastre and Mapping Agency (NCMA). This is its main interface.**

Emotional Stability significantly affects the type of reported usability issues. Problem severity is not affected by any trait. Valence ratings are significantly affected by Conscientiousness, whereas Agreeableness, Emotional Stability and Openness significantly affect arousal ratings. Finally, Openness has a significant effect on the number of detected peaks in user's skin conductance.

## 1 INTRODUCTION

User experience (UX) emerged as a new research area emphasizing in qualitative aspects of user interaction [2,11]. Beyond collecting traditional metrics (e.g. task success rate, time on task), emotional assessment is the main aspect of UX evaluation [2,19]. There are subjective [12,21] and objective [14,16] approaches to measure emotions in UX evaluation. One popular tool for subjective emotional assessment is the two-dimensional (2D) Affect Grid tool [17] (Fig.1). Collecting and analyzing data from users' physiology (e.g. heart rate, skin conductance) is an objective approach of emotional assessment [15,22] and has recently gained much attention. Skin conductance is particularly sensitive to emotional fluctuation [10,14,15] and was measured in this study.

In the context of UX evaluation, it is important to highlight any individual aspect that may affect results. Research in psychology has revealed significant effects of personality traits on individual behavior. For instance, personality has been found to be a reliable predictor of participants' learning style [23] and creativity [3]. Studies have also shown that personality traits affect the way users accept and interact with technology [7,8]. For instance, it is well-known that there are differences on how people feel and how they rate the usability of a product while interacting with it. Such discrepancies may exist due to distinct personality traits [7].

Although users' selection criteria, such as level of experience (novice vs. expert users), demographics (e.g. age, gender) and cultural background have been investigated [5,13,18,20], the effect of participants' personality characteristics on UX evaluation metrics has been seldom explored. Studies that address this issue, [1,4] either focus on the analysis of a specific personality trait (e.g. extroverts vs introverts) or on a specific target group (e.g., children). In specific, the present study addresses the following research questions:

- RQ1: Do personality traits affect the number, type and severity of identified usability issues?
- RQ2: Do personality traits affect users' emotional assessments as operationalized by Valence-Arousal ratings?
- RQ3: Do personality traits affect participants' stress level as measured by fluctuation in participants' skin conductance?

## 2 METHODOLOGY

### 2.1 Participants and Procedures

Twenty-four typical computer users (14 males), aged between 18 and 45 (M=32.3, SD=7.5) were recruited from the University campus. Participants were students and administrative staff.

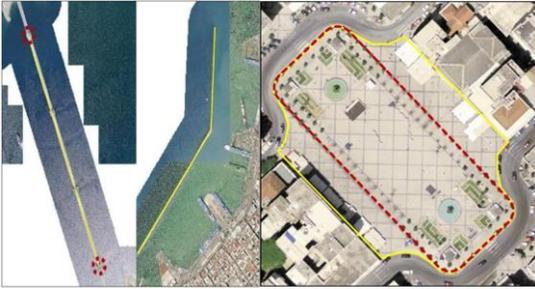

**Figure 3: Scenario 1 (left) involved measuring the distance between the first and the fourth pillar of a well-known bridge and the breakwater's length in the old harbor. Scenario 2 (right) involved measuring the inner area of a popular square located in the city's center (red-dotted rectangle) and modifying this area to include the side parts of the square (yellow polygon).**

**Table 1: Nielsen's severity rating for usability problems [24]**

| Rating | Description |
|---|---|
| 1 | Cosmetic issue |
| 2 | Minor issue |
| 3 | Major issue |
| 4 | Usability catastrophe |

**Table 2: Median Score per Personality Trait**

| Personality trait | Median |
|---|---|
| Extraversion: (Trait1) | 28.0 |
| Agreeableness: (Trait2) | 41.0 |
| Conscientiousness: (Trait3) | 36.5 |
| Emotional stability: (Trait4) | 29.0 |
| Openness to exp.: (Trait5) | 33.5 |

Participants were asked to interact with the web-based service, offered by the National Cadastre and Mapping Agency (NCMA) (Fig.2) to perform two tasks (Fig.3). This service was selected because a previous pilot user testing study with 5 participants found it unusable (e.g. none of the participants managed to complete a task). None of the participants had previous experience with the service. The tasks were designed to include well-known places to control for the effect of spatial knowledge. Next, they were asked to complete a consent form, questions for demographics and the 50-item Big-five Trait Test. Afterwards, the skin conductance sensor was placed on them. The interaction scenarios were presented to participants in a counterbalance mode to remove potential confounds created by task presentation order. Finally, users were engaged in a Retrospective Think Aloud (RTA) session just after interacting with the system.

### 2.2 Metrics and Instruments

*2.2.1 Personality Traits.* The 50-item Big-five Trait Test questionnaire was provided in participant's native language.

*2.2.2 Usability Issue-based Metrics.* Each usability issue was noted using a user id, issue id, and a description given by the user during RTA session (e.g. "I couldn't find the appropriate tool"). Five HCI experts (two Professors of HCI, two postdoctoral HCI researchers, and one doctoral student of HCI, all with at least 5 years of professional experience in UX evaluation) were provided with the list of all the identified usability issues, tasks' descriptions and participants' screen recordings.

Next, they were asked to assign the user-reported issues to types of problems based on Nielsen's 10 heuristic rules [25] and rate the severity (Table 1) of each one by taking into account the frequency, impact of occurrence and persistence of the problem [24]. They worked independently and then met to finalize their assignments following a consensus-based decision-making process as it is typically done in evaluation settings involving experts.

*2.2.3 Emotional Ratings.* During the RTA session each participant rated their own reported usability issues in the emotional scale of Valence (from 1 to 9)–Arousal (from 1 to 9).

*2.2.4 Physiology-based Metrics.* Skin conductance was recorded with a sampling rate of 32Hz using the NeXus-10 physiological platform.

## 3 RESULTS AND ANALYSIS

Our dataset includes 116 usability issues with an associated participant's emotional (Valence-Arousal) rating. Each issue has also an associated severity rating provided by the five HCI experts. Regarding physiological signals, a mean number of 10 significant peaks (SD=6.8) in participants' skin conductance was counted using PhysiOBS, a UX data analysis tool that, inter alia, supports automated signal smoothing and segmentation based on significant peaks (see [14] for details).

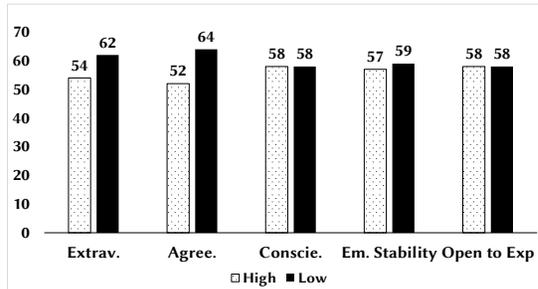

**Figure 4: Distribution of reported issues in each trait per group (High-Low) after median split (Table 2).**

**Table 3: Descriptives for the Number of Identified Issues and Severity per Personality Trait Group**

| Personality Trait | Group | Mean Issues | SD | Mean Severity | SD |
|---|---|---|---|---|---|
| Extraversion | High | 5.67 | 1.67 | 2.58 | 0.78 |
| Extraversion | Low | 4.25 | 1.77 | 2.63 | 0.81 |
| Agreeableness | High | 4.73 | 1.90 | 2.69 | 0.84 |
| Agreeableness | Low | 5.15 | 1.82 | 2.54 | 0.76 |
| Conscient. | High | 4.92 | 1.88 | 2.70 | 0.76 |
| Conscient. | Low | 5.00 | 1.86 | 2.51 | 0.83 |
| Emo. Stability | High | 5.58 | 2.15 | 2.64 | 0.75 |
| Emo. Stability | Low | 4.33 | 1.23 | 2.57 | 0.84 |
| Open to exp. | High | 4.10 | 0.99 | 2.70 | 0.87 |
| Open to exp. | Low | 5.57 | 2.07 | 2.52 | 0.71 |

The collected data were organized into groups per personality trait based on a median-split of the associated Big-five score (Fig.4). In all subsequent statistical analyses, effect sizes were calculated using the formulas found in [9] and are reported only in cases of significance. In addition, data normality and homogeneity of variance assumptions were tested using Shapiro-Wilk and Levene's test respectively.

### 3.1 RQ1: Effect of Personality Traits on Usability Issue-based Metrics

*3.1.1 Personality Traits and Number of Usability Issues.* In the Openness trait, an independent samples t-test found a significant difference for the reported usability issues between user groups (Table 3); t(22)=2.08, p=0.031, r=0.46. This medium-to-large observed effect size [6] demonstrates the importance of the Openness trait on participants' effectiveness in identifying usability issues during RTA: users scoring lower in Openness find significantly more issues. The rest four traits did not significantly affect the number of identified usability issues in RTA.

*3.1.2 Personality Traits and Types of Usability Issues.* A significant difference was observed only for the problems grouped in the heuristic "Visibility of system status". In specific, Mann-Whitney analysis found that users scoring higher in the Emotional Stability trait reported significantly more usability issues violating this heuristic (Mdn=3.58) compared to the ones scoring lower in this trait (Mdn=2.50); Z=1.98, p=0.047, r=0.41. Again, a medium-to-large effect size [6] was observed.

*3.1.3 Personality Traits and Severity of Usability Issues.* No significant difference was observed between the severity of usability problems found in the low and high groups of each trait (Table 3).

### 3.2 RQ2: Effect of Personality Traits on Emotional Assessment (Valence-Arousal)

*3.2.1 Personality Traits and Valence Ratings.* A Mann-Whitney test showed that participants' valence ratings were significantly higher for the low-group compared to the high-group in the Conscientiousness trait (Table 4); Z=1.96, p=0.049, r=0.18.

*3.2.2 Personality Traits and Arousal Ratings.* A Mann-Whitney test indicated that the arousal ratings were significantly higher for the low-group than for the high-group in both the Agreeableness and Openness traits (Table 4); Z=2.55, p=0.011, r=0.24 and Z=3.60, p=0.001, r=0.34 respectively. By contrast, in the Emotional Stability trait Mann-Whitney showed that arousal ratings were significantly higher for participants in the high-group condition than in the low-group condition (Table 4); Z=2.28, p=0.022, r=0.22.

### 3.3 RQ3: Effect of Personality Traits on Skin Conductance

Regarding participants' skin conductance, the only significant difference was found in the Openness trait (Table 5). In specific, users' skin conductance peaks were significantly less in the high-group compared to the low-group Openness condition; t(22)=2.44, p=0.023, r=0.46. This is also a medium-to-large observed effect size [6].

Table 4: Descriptives for the Valence (V) & Arousal (A) Rating per Personality Trait Group

| Personality Trait | Group | Mean (V) | SD (V) | Mean (A) | SD (A) |
|---|---|---|---|---|---|
| Extraversion | High | 3.42 | 1.07 | 5.21 | 1.92 |
| Extraversion | Low | 3.06 | 1.08 | 5.47 | 1.89 |
| Agreeableness | High | 3.12 | 1.09 | 4.86 | 1.90 |
| Agreeableness | Low | 3.32 | 1.09 | 5.75 | 1.82 |
| Conscient. | High | 3.04 | 1.18 | 5.40 | 1.98 |
| Conscient. | Low | 3.42 | 0.96 | 5.30 | 1.83 |
| Emo. Stability | High | 3.09 | 1.08 | 5.77 | 1.88 |
| Emo. Stability | Low | 3.36 | 1.09 | 4.95 | 1.85 |
| Open to exp. | High | 3.21 | 1.00 | 4.70 | 1.80 |
| Open to exp. | Low | 3.24 | 1.17 | 5.98 | 1.79 |

Table 5: Descriptives for the Skin Conductance Peaks per Personality Trait Group

| Personality Trait | Group | Mean | SD |
|---|---|---|---|
| Extraversion | High | 10.33 | 8.21 |
| Extraversion | Low | 9.58 | 5.49 |
| Agreeableness | High | 10.64 | 7.00 |
| Agreeableness | Low | 9.38 | 6.92 |
| Conscient. | High | 11.67 | 5.23 |
| Conscient. | Low | 8.25 | 8.00 |
| Emo. Stability | High | 10.83 | 6.95 |
| Emo. Stability | Low | 9.08 | 6.91 |
| Open to exp. | High | 6.30 | 4.97 |
| Open to exp. | Low | 12.57 | 6.93 |

## 4 CONCLUSIONS, LIMITATIONS AND FUTURE WORK

This study showed that users less Open to Experience report significantly more usability issues in RTA. In addition, users scoring higher in Emotional Stability report significantly more violations of the "Visibility of system status" problem type. Personality traits do not affect the severity of reported usability issues in an RTA session. Furthermore, it was found that users scoring higher in the Conscientiousness trait provide significantly lower Valence ratings. In addition, people that are more Agreeable and Open to Experience rate their arousal significantly lower. By contrast, people scoring higher in Emotional Stability rate higher their arousal. Skin conductance analysis revealed that participants who are more Open to Experience have significantly lower stress while confronting usability issues.

One particularly interesting pattern is that the less Open to Experience participants are, the more they experience (skin conductance peaks) and report being stressed (valence-arousal ratings), and also report more usability issues in RTA. This is work-in-progress research but it might have practical implications for the efficiency of users' screening process. For instance, if one is mostly interested in identifying the most stressful usability issues, then a smaller number of potential participants with the lowest Openness to Experience personality trait score might be adequate.

The research presented in this paper is not without limitations. First, all issue-based dependent variables (count, type, severity) are associated with the application of the RTA protocol. It remains unclear if the findings hold true when an alternative protocol is applied, such as concurrent thinking aloud, or even when a different system is evaluated. Additional studies are required and constitute one of our immediate future research goals. One additional limitation concerns the effect of personality traits on usability issue type. In our study, there were heuristics that were assigned with a low number of usability issues (e.g., Heuristic 10: Help and documentation). In addition, alternative approaches could have been employed to group usability issues into categories, such as some other taxonomy of usability problems or even thematic analysis of the problems found. These factors might affect the findings and can be the object of future research.